\documentstyle[art12]{article}
\begin{document}
\rightline{INP MSU preprint-96-33/440}
\rightline{(revised version)}
\def\emline#1#2#3#4#5#6{%
       \put(#1,#2){\special{em:moveto}}%
       \put(#4,#5){\special{em:lineto}}}
\begin{center}
  \begin{Large}
  \begin{bf}
DIFFRACTIVE PHOTOPRODUCTION OF MESON PAIRS IN THE S\"ODING MODEL \\
  \end{bf}
  \end{Large}
  \vspace{5mm}
  \begin{large}
    N.I. Starkov\\
  \end{large}
P.N. Lebedev Physical Institute, Russian Academy Sciences, \\
Moscow 117924, Russia\\
  \vspace{5mm}
  \begin{large}
    N.P. Zotov\\
  \end{large}
D.V. Skobeltsyn Institute of Nuclear Physics, \\
M.V. Lomonosov Moscow State University,
      Moscow 119899, Russia\\
\end{center}
  \vspace{10mm}
\begin{abstract}
We consider the diffractive photoproduction of $\pi^+\pi^-, K^+K^-,
D^+D^-$ meson pairs with high masses and transverse momenta in 
the framework of the S\"oding model.
Total and differential cross sections are calculated for these processes.
The role of form factors is stressed.
\end{abstract}

\section{Introduction}
 
 The experimental results obtained  by H1 and ZEUS Collaborations on 
HERA ep-collider for diffractive $\rho^o, \phi$, and $J/\Psi$ meson production 
by real and virtual photons attract abnormally high attention~\cite{r1}.
 The main reason of this attention is explained by the fact
that transition from small virtuality photon $(Q^2\simeq 0)$ to deep inelastic 
scattering $(Q^2 > 4$ GeV$^2$) connects with dynamic of a transition from soft 
diffractive $\gamma p$ interaction to hard one which is described by the vacuum
number exchange, the so called Pomeron~\cite{r2}.

Recently H1 and ZEUS Collaborations published experimental data on diffractive
$\rho^0$  meson photoproduction~\cite{r3,r4}. These results confirmed  the
general properties of soft diffractive hadron processes which are well known, but
also showed the necessity for further experimental and theoretical studies.
 
It is known that the cross section for diffractive dissociation of an incoming
 particle (a photon
on our case) into low mass states contains at all three different contributions
(see for instance~\cite{r5}): 1) the term describing the direct production
 of hadron states (resonances)
which then decay into final state hadrons, 2) the Drell--Hiida--Deck (DHD) nonresonant
background term (more precisely the resonance like one)~\cite{r6}, and
3) the terms originating  from final state interactions~\cite{r7}.

The DHD mechanism for the nonresonant background in $\rho^0$ meson
 photoproduction has been used
by S\"oding ~\cite{r8}. The interference between the amplitude of 
DHD nonresonant $\gamma p\to
\pi^+\pi^-p$ production and the amplitude of resonant $\gamma 
p\to\rho^0p\to\pi^+\pi^-p$ production
leads to a shift of $\rho$ the mass spectrum. In general the correct
account of the interference pion pairs produced by  the DHD nonresonant mechanism and
via the $\rho^0$ meson decay is a theoretical as well as an experimental problems 
~\cite{r5,r7}.

Therefore in this note we propose a simple model for nonresonant diffractive meson
pair ($\pi^+\pi^-, K^+K^-, D^+D^-$) production which contains no diffractive resonance
($\rho^0, \phi,$...) production.

The paper is organized as follows. In Section 2 we formulate the model (similar to the
S\"oding model) for the meson pair photoproduction. In Section 3 we present our numerical
results for diffractive photoproduction of meson pairs. Section 4
devotes qualitive study of the consequences of the absorption
in the final state on  $t$ distribution of meson pairs.
In Conclusions we emphasize
an importance of the proposed model for a description of many background processes
in a number of experiments on HERA collider.

\section{A model for diffractive photoproduction of meson pairs} 

We use S\"oding model~\cite{r8} for diffractive nonresonant photoproduction of meson
pairs: $\pi^+\pi^-, K^+K^-,$ and $D^+D^-$. In according with this model a
photoproiduction of meson pairs is described by two diagrams containing 
the exchange of a Pomeron
(Fig.1), which correspond to the following formula for the differential cross section
for meson pair production by transversely polarized photons ($Q^2 = 0$):
\begin{equation}
\frac{d\sigma^T}{dt}=\frac{\alpha}{2\pi^2}e^{Bt}\frac{\sigma^2_{tot}}{16\pi}
\int\frac{z(1-z)dzd^2{q}_{t(-)}}{m^2_{(+)t}m^2_{(-)t}}AF^2(q^2_{t(-)}),
\end{equation}
where $\sigma_{tot}$ and $B$ are the usual parameters describing meson--proton 
scattering: the total cross section and the slope parameter of diffractive cone.
The two--component transverse momentum of a meson $\vec{q}_{t(-)}$ is defined in the
$\gamma p$ c.m.s. using the Sudakov decomposition~\cite{r9}
for 4--momentum of the negative charge meson in final state: $q_{(-)} = zQ + \beta p +
q_{t(-)}$,
with $z = (q_{(-)}p)/(Qp)$ and $\beta = m^2_{(-)t}/[z2(Qp)]$.
Here $Q$ and $p$ are the 4--momenta of the photon and the proton (with $m_p$ = 0),
respectively. The factor $A$ in (1)
has the following form \footnote{
The approximate expression for $A$ initially was obtained by M.Ryskin~\cite{r10}}:
\begin{equation}
A = \left[\frac{(1-z)}{z}\frac{m^2_{(-)t}}{m^2_{(+)t}}q^2_{t(+)} + \frac{z}{(1-z)}
\frac{m^2_{(+)t}}{m^2_{(-)t}}q^2_{t(-)} - 2\vec{q}_{t(+)}\vec{q}_{t(-)})\right]
\end{equation}
and is equal to $\vec{q}\,^2_{t(-)}/[z(1-z)]$ at $t = 0$. The remaining kinematic variables in
(1) and (2) have the following form: $m^2_{(-)t} = \mu^2 + \vec{q}\,^2_{t(-)}$ 
is transverse
mass of a meson, $t\simeq -(\vec{p}\,'_t)^2$ and $\vec{q}_{t(+)} = -\vec{p}\,'_t -
\vec{q}_{t(-)}$.

In (1) the functions $F(q^2_{t(-)})$ are "form factors" of the two meson states 
connected by off mass shell mesons.\footnote{In the original S\"oding model
~\cite{r8} these form factors are absent.} 
The exact expressions for these form factors are unknown.  
They can be written, for instance, in pole
forms, where they are expressed through  $\vec{q}\,^2_{t(-)}$ toghether
with masses of suitable
meson states. For $\pi, K,$ and $D$ mesons these masses can be chosen to be
the masses of $\rho, \phi$, and $J/\Psi$ mesons (in the following  named "version 1")
or of $\pi(1300), \phi$, and $J/\Psi$ mesons~\cite{r10} ("version 2").
 We use here both
possibilites. As an example, for "version 2" the form factors for $\pi$ pair
photoproduction is 
\begin{equation}
F_{\pi}(q^2_{t(-)}) = \frac{m^2_{\pi(1300)}}{m^2_{\pi(1300)} + 
 \vec{q}\,^2_{t(-)}}.
\end{equation}
In the following, the total cross sections for $\pi p$ and $Kp$ scatterings in (1)
are assumed to be described by 
the Donnachie - Landshoff parametrizations~\cite{r11}, which give at HERA energies 
($\sqrt{s} = 100 GeV$) the following values: 
$\sigma_{tot}(\pi p) = 29.2\, mb,\,\, \sigma_{tot}(Kp) = 25.1\, mb$. For
$\sigma_{tot}(Dp)$ we used the different values ranging from $3\, mb$ ("version 1")
up to $\sigma_{tot}(Dp)\,=\,\sigma_{tot}(\pi p)/2$ ("version 2").
The values of $B-$ parameters in (1) 
are equal: $B_{\pi p} = B_{Kp} = 10\, GeV^{-2}$ and $B_{Dp} =
6\, GeV^{-2}$.

We would like to point out that the values of the cross section $\sigma_{tot}(Dp)$ and 
the slope parameter $B_{Dp}$ are not determined enough exactly so far and the obtained
results for the $D $ meson photoproduction cross sections depend essentially 
on these ones and  corresponding form factor.
 
 The distribution $d\sigma^T/dM$ over the mass of a meson pair, 
$M = (q_{(+)} + q_{(-)})$, 
can be obtained through the transition from $z$ to $M$ in eq. (1).
We used the following relations valid at $t = 0$:
\begin{equation}
z(1-z)M^2 = \mu^2 +  \vec{q}\,^2_{t(-)},\,\,  dz/dM =2Mm^2_t/[M^4(1-4m^2_t/M^2)]
\end{equation}
 
\section{Numerical results} 

 The differential cross section $d\sigma/dt$ for the processes  of diffractive 
photoproduction of meson pairs ($ \pi^+\pi^- $, $K^+K^- $ , and $ D^+D^-)$  are 
presented in Fig.2. 
The slopes of these cross sections are determined in the main by $B$ parameters in (1).

We evaluated the values of these differential cross sections at $t = 0$. Based on the form
factors of "version 1" we obtained
45.3 $\mu b/GeV^2 (\pi^+\pi^-$ pairs), 14.5 $\mu b/GeV^2 (K^+K^-$ pairs) and 0.14
$\mu b/GeV^2 (D^+D^-$ pairs). "Version 2" resulted in 
90 $\mu b/GeV^2 $, 14.5 $\mu b/GeV^2 $ and 3.5 $\mu b/GeV^2$ correspondingly.


The transverse momentum distributions of outcoming mesons 
$d\sigma /dt dq^2_{t(-)}$ for the process 
$\gamma p \to K^+K^-p$ at a fixed values $|t|\,=\,0$ and $2\,(GeV/c)^2$
are shown 
in Fig. 3.
These
distributions at $|t|=0$ and/or $|t| \ne 0$ but at high $|\vec{q}_{t(-)}| > 3 
GeV/c$ have a behaviour $\sim |\vec{q}_{t(-)}|^{-6}$.

The cross sections $\sigma (|\vec{q}_{t(-)}| > |\vec{q}_{t(-)}|_{min})$ for
$\gamma p \to \pi^+\pi^- p ,\, \to K^+K^- p$, and $\to D^+D^- p$ processes 
shown in Fig.4.
The  total cross section for diffractive meson pair photoproduction was obtained
by integrating expression (1). Using the form factors of "version 1" results in
4.6  $\mu b$ for $\pi^+ \pi^-$ pairs, 1.4  $\mu b$
for $K^+ K^-$ pairs and $\sim$ 0.02 $\mu b$ for $D^+D^-$ pairs.
Based on the form factors of "version 2" we obtained:
10 $\mu b$, 1.4 $\mu b$
and $\sim$ 0.5 $\mu b$ correspondingly.


The results for the mass distributions of  diffractive photoproduction of pion
and kaon pairs are shown in Fig.5. The curves (1) and (2) for the pion pair
photoproduction correspond to two versions of the form factor (3). The
value of the background in the $\rho^0$ mass region $0.55 < M < 1\, GeV$ 
somewhat larger than the ZEUS estimates\cite{r4} and  at larger
mass there is  a strong dependence on the choice of the form factor.

The total cross sections obtained by integrating $d\sigma/dM$ over the mass $M$
of a meson pairs are presented in Fig.6. The values of these cross sections
at high $M$ coincide approximately with those obtained by integrating expression (1)
over $z$ and $q_{t(-)}$.

As we have pointed in Introduction meson pair photoproduction cross
sections contain final state interaction contribution besides
resonant and non-resonant ones. The effects of rescattering in final 
state (or final state absorption) for $\pi^+\pi^-$ - photoproduction
can lead to 20-40\verb!%! corrections~\cite{r5,r12}. But for
photoproduction of heavy mesons ($K$ and $D$) our model is more adequate.

\section{Some remarks on absorptive corrections}

 The expression for differential cross section (1) takes into account 
 the contribution of the elastic meson - proton scattering and also
all the rescatterings of one meson on proton (see Fig.1). Then for the elastic
(diffractive) photoproduction of meson pairs one should iclude two kinds of the 
absorptive corrections in the final state: the elastic rescattering of mesons on each
other and one of both mesons off the target (proton). The former corrections are
essential for the resonance (for instance $\rho $ meson) production of meson
pairs can be take into account by the special prescription~\cite{r7,r12}
and do not affect considerably on the slope parameter of differential cross
section.

The things  go quite differently for the absorptive corrections connected
with the rescattering of two meson system  as whole off the target. These corrections
lead to strong effects for t-dependence of meson pair differential cross
section (see, for instance~\cite{r5}).

Here we restrict ourselves to the allowance of these 
corrections on the phenomenological level,
when in the eikonal approximation the dependence $d\sigma/dt\sim\exp{Bt}$ is replaced 
by~\cite{r12}:
\begin{equation}
   \frac{d\sigma}{dt}\sim \left [\exp{(Bt/2)} - \frac{\sigma}{16\pi B}\exp{(Bt/4)}
\right ]^2
\end{equation} 

The absorptive corrections in final state may be enhanced due to  inelastic
diffractive intermediate states in the proton vertex of diagram in Fig. 1, then
$\sigma$ is replaced by $\lambda\sigma$~\cite{r5,r13} in the second term of eq. (5).
The phenomenological parameter $\lambda$ should have the following value:
$\lambda \approx 1 + \sigma_{D}/\sigma_{el}$~\cite{r14}. Some estimates for 
$NN -$interactions give $\lambda \approx 1.2$~\cite{r15}; in the region of the
laboratory energy $E_L = 10 - 10^3\, GeV/c$ the ratio $\sigma_D/\sigma_{el}\approx 1$
and $\lambda \approx 2$. Therefore in our calculations we used the following values:
$1.2 < \lambda < 2$~\cite{r5}.

The results of our calculations of differential cross section of the process
$\gamma p \to \pi^+\pi^- p$ by the formula (5) are ilustrated in Fig. 7
at suitable set of parameters (for arbitrary normalization of $d\sigma/dt):
\sigma = 28\, mb, B = 7\,(GeV/c)^{-2}$ and $\lambda = 1.3$.

In rough approximation for the allowance of the absorbtive corrections 
differential cross section $d\sigma/dt$ has a dip at $t\simeq 0.75\, (GeV/c)^2$,
which may be filled in a more ralistic model. But the main point is that the
$d\sigma/dt$  changes of the slope parameter at high $t$ (at $\mid t\mid > 0.8 \,(GeV/c)^2$),
which is observable in the preliminary ZEUS experimental data for $\gamma p \to
p\pi^+\pi^-$ at $M_{\pi^+\pi^-} = \,0.5 - 1.1\, GeV$ in the region
$W = 64 - 90\, GeV$ and $0.25 < Q^2 < 0.8\, GeV^2$~\cite{r16}. Similar dips
may appear in differential cross sections of diffractive $J/\Psi$ and
$\rho^0 -$productions in DIS~\cite{r17,r18}.  

\section{Conclusions}

  We presented a new model for diffractive meson pair photoproduction which we
applied to estimate the non-resonant meson pair photoproduction
cross sections.
We showed
that the results strongly depend on the choice of form factors.
The $t -$dependences of differential cross sections are determined
essentially by the absorptive corrections  in  final state.
Although the non-resonant cross-sections are small in comparision 
with the resonant diffractive meson pair photoproduction, they
strongly affect background processes in $\rho^o , \phi $ and
$J/\psi$  meson production by real and virtual photons.

First of all the known mass shift effect for $\rho$ - meson 
photoproduction which is controled ~\cite{r8} by the S\"oding mechanism
of the non-resonant $\pi^+\pi^-$  pair photoproduction ~\cite{r3,r4}.
The S\"oding mechanism contribution is visible at low $Q^2\ne0$ 
\footnote{We are grateful to H. Beier for this remark.} and it can be
estimate in the framework of our model (see also~\cite{r19}).
 
The second our estimations for non-resonant diffractive $K^+K^-$ 
meson photoproduction contribution are important for correct 
extraction of $K^-$  meson photoproduction cross section at
HERA energies.  As was noted by Pumplin
~\cite{r12} the process $\gamma p \to K^+K^- p$ is interesting
because it could be studied at low $K^+K^-$ masses without interference
from $\phi-$resonance production.

Our proposed $D^+$  meson photoproduction mechanism should give 
contribution to the $\mu$ - production at large transverse momentum
$q_t$. These muons can give large background for muon pairs from
$J/\psi$ meson production on HERA collider.

Finnally the formula (1) can be used for extraction of the elastic
$\pi^\pm p $, $K^\pm p $ and $D^\pm p $ cross sections~\cite{r19} \footnote{
Initially it was proposed by M. Arneodo} if we have
experimental values for non-resonant diffractive 
$\pi^+\pi^-$, $K^+K^- $ and $D^+D^-$ pair photoproduction cross
sections.
\newpage
\section{Acknowledgements}

One of us (N.Z.) would like to thank M. Ryskin for the original suggestion to study 
diffractive non-resonant meson pair photoproduction and for
useful disscussions, A. Proskuryakov for discussions  of correlations
of the resonant and background contributions in meson pair photoproduction and
J. Repond for the careful reading of the manuscript and useful ramarks.

{\bf Figure captions}
\begin{enumerate}
\item
Diagrams for diffractive meson pair photoproduction in S\"oding model.
\item
The differential cross sections $d\sigma/dt$ for diffractive meson pair
 photoproduction.      
\item
The transverse momentum distributions for $\gamma p \to K^+K^- p$ at fixed values
of $|t|$.
\item
The integral cross sections for diffractive meson pair photoproduction
processes as a function of $q_{min}$.
\item 
The mass distribution $d\sigma/dM$ over the mass of two $\pi -$ and $K -$mesons:
the curves (1) and (2) for the version 1 and 2 of form factors.
\item
The total cross sections $\sigma (M_{max})$ versus the mass of two $\pi -$
and $K -$mesons: curves as in Fig.5.
\item
The differential cross section for the process  $\gamma p \to \pi^+\pi^- p$
obtained by the formula (5).
\end{enumerate}
\end{document}